# Affinity, stoichiometry and cooperativity of heterochromatin protein 1 (HP1) binding to nucleosomal arrays

**Vladimir B. Teif**[1,*], **Nick Kepper**[1], **Klaus Yserentant**[1,2], **Gero Wedemann**[3] & **Karsten Rippe**[1,*]

[1] Deutsches Krebsforschungszentrum & BioQuant, Im Neuenheimer Feld 267, 69120 Heidelberg, Germany

[2] present address: Cell Networks Cluster & Institut für Physikalische Chemie, Universität Heidelberg, Im Neuenheimer Feld 267, 69120 Heidelberg

[3] CC Bioinformatics, University of Applied Sciences Stralsund, Zur Schwedenschanze 15, 18435 Stralsund, Germany

*E-mails: v.teif@dkfz.de, karsten.rippe@dkfz.de

**Abstract.** Heterochromatin protein 1 (HP1) participates in establishing and maintaining heterochromatin via its histone modification dependent chromatin interactions. In recent papers HP1 binding to nucleosomal arrays was measured *in vitro* and interpreted in terms of nearest-neighbor cooperative binding. This mode of chromatin interactions could lead to spreading of HP1 along the nucleosome chain. Here, we reanalyzed previous data by representing the nucleosome chain as a one-dimensional binding lattice, and show how the experimental HP1 binding isotherms can be explained by a simpler model without cooperative interactions between neighboring HP1 dimers. Based on these calculations and spatial models of dinucleosomes and nucleosome chains, we propose that binding stoichiometry is dependent of the nucleosome repeat length (NRL) rather than protein interactions between HP1 dimers. According to our calculations, more open nucleosome arrays with long DNA linkers are characterized by a larger number of binding sites in comparison to chains with short NRL. Furthermore, we demonstrate by Monte Carlo simulations that the NRL dependent folding of the nucleosome chain can induce allosteric changes of HP1 binding sites. Thus, HP1 chromatin interactions can be modulated by the change of binding stoichiometry and type of binding to condensed (methylated) and non-condensed (unmethylated) nucleosome arrays in the absence of direct interactions between HP1 dimers.

**Introduction**
The eukaryotic genome is organized inside the cell nucleus by histones and other chromosomal proteins into a nucleo-protein complex called chromatin. At the first level of compaction, 145-147 DNA base pairs (bp) are wrapped around the histone octamer to form the nucleosome – the elementary chromatin unit (Olins and Olins, 1974; Kornberg, 1974; van Holde, 1989). At the next level, nucleosomes connected by the linker DNA are organized via their interaction into more compacted chains (Schiessel, 2003; Belmont, 2014; Hansen, 2002; van Holde, 1989; Woodcock, 2006; Robinson and Rhodes, 2006). Nucleosome fibers can have different structure and properties depending on the their spatial organization and epigenetic modification, which in turn determines access to the DNA and gene expression levels. In a simplified view, two major types of chromatin states can be distinguished: heterochromatin as a more compact and less transcribed conformation, in contrast to the more open euchromatin, which is enriched in actively transcribed genes. In a more detailed classification, multiple chromatin states can be identified as for example by principal component analysis according to the chromosomal protein content (Filion *et al.*, 2010). One example of a chromatin protein that demarcates silenced chromatin is heterochromatin protein 1 (HP1) with the yeast homologue Swi6 (Hiragami and Festenstein, 2005; Kwon and Workman, 2008; Maison and Almouzni, 2004). The N-terminal chromodomain (CD) and the C-terminal chromoshadow-domain (CSD) of HP1/Swi6 are connected by a flexible linker region. The CD interacts specifically with H3 histone tails that are trimethylated at lysine 9 (H3K9me3) (Fischle *et al.*, 2005; Jacobs *et al.*, 2004), while the CSD promotes the formation of HP1 dimers (Brasher et al., 2000; Nielsen et al., 2001; Yamamoto and Sonoda, 2003; Rosnoblet et al., 2011). Since the interaction of HP1 with the histone methylase Suv39h (Clr4 in yeast) via the CSD has been demonstrated, HP1 binding-driven propagation of H3K9me3 and HP1 has been proposed (Grewal and Jia, 2007; Eissenberg and Reuter, 2009; Schotta et al., 2002). Recent studies investigated the H3K9me3 dependent binding of the yeast HP1 homologue Swi6 to different nucleosome substrates in vitro (Canzio *et al.*, 2011; Al-Sady *et al.*, 2013; Canzio *et al.*, 2013; Canzio *et al.*, 2014). From these studies it was concluded that cooperative binding and oligomerization and spreading of Swi6 on chromatin can contribute to propagating the heterochromatin state along the nucleosome chain. While a number of theoretical models have been developed to account for the switching and propagation of nucleosome states via this type of feedback loop mechanisms (Stein, 1980; Hong *et al.*, 2011; Micheelsen *et al.*, 2010; Sneppen *et al.*, 2008; Dodd *et al.*, 2007; David-Rus *et al.*, 2009; Sedighi and Sengupta, 2007; Angel *et al.*, 2011; Zhao-Wen *et al.*, 2011), the mechanistic details of this process remain far from being understood. There are also specific theoretical issues that need to be considered. For example, long-range but not contact interactions between DNA-bound proteins can lead to the phase transition of the first kind due to protein-DNA binding (Lando and Teif, 2000; Teif *et al.*, 2002). A series of later works applied similar ideas of long-range interactions to the spreading of the heterochromatin state (Micheelsen *et al.*, 2010; Sneppen *et al.*, 2008; Dodd *et al.*, 2007). Yet, long-range interactions up to now have no experimental support for the case of HP1 chromatin binding, while contact HP1-HP1 interactions would not be sufficient to explain a heterochromatin-euchromatin boundary in the absence of additional factors like insulator or DNA sequence-determined modification domains.

Most of the theoretical approaches consider the DNA as a 1D lattice of units, which can adopt different states depending on reversible binding or covalent modifications (Chereji and Morozov, 2014; Vaillant *et al.*, 2010; Schwab *et al.*, 2008; Mobius *et al.*, 2013; Morozov *et al.*, 2009; Segal *et al.*, 2006; Kaplan *et al.*, 2009; Teif and Rippe, 2011; Teif and Rippe, 2010; Teif *et al.*, 2010; Wang *et al.*, 2008; Lubliner and Segal, 2009; Florescu *et al.*, 2012; Teif and Rippe, 2009; Afek *et al.*, 2011; Mirny, 2010; Gabdank *et al.*, 2010; Chevereau *et al.*, 2009; Reynolds *et al.*, 2010). The size of the elementary lattice unit can be set either as one DNA base pair, or one nucleosome, or one can use an even larger coarse graining. Thus, these models can be scaled to study genomic events both at the molecular and the systems level (Teif *et al.*, 2013). The major challenge, however, is the parameterization of the model based on the experimental data as these determine principally different properties of the model.

A series of recent studies provided valuable insights into the nucleosome binding properties of HP1/Swi6 (Canzio *et al.*, 2011; Canzio *et al.*, 2014; Canzio *et al.*, 2013). Swi6 was found to bind nucleosomes as a dimer with a stoichiometry of two Swi6 dimers per mononucleosome, and assembled on the nucleosomal lattice in a manner strongly affected by the H3K9 methylation state. From an analysis of the binding curves it was concluded that attractive interactions between neighboring Swi6 dimers exist that lead to spreading of Swi6 along the nucleosome chain. Furthermore, the length of the DNA linker between nucleosomes had a large effect on HP1 binding. In order to further dissect the mode of Swi6 binding to chromatin, we have performed here calculations using an analytical 1D lattice binding algorithm and 3D Monte Carlo simulations of nucleosome chain folding. The results prompt us to propose an alternative model for Swi6 binding that considers changes of nucleosome chain folding depending on the nucleosome repeat length.

**Model**
In the previous experiments, Swi6 binding to mono- and di-nucleosomes and 12-mer nucleosomal arrays was quantified by measuring the fraction of nucleosomal substrate without any bound Swi6 by gel electrophoresis (Canzio *et al.*, 2011). Thus, configurations with one or more bound Swi6 proteins per nucleosomal substrate were not distinguished. This is different from the conventional representation of the degree of binding as the fraction of occupied lattice binding sites. In particular, the concentration of half-saturation of such titration curves is not directly related to the standard dissociation constant for protein binding to a nucleosomal lattice. This warrants additional consideration for dissecting binding affinity and cooperativity parameters. In a statistical-mechanical lattice binding formalism, the unbound fraction of nucleosome arrays can be simply expressed as $1/Z$, where $Z$ is the partition function of the system (Teif *et al.*, 2010; Teif and Rippe, 2010; Teif and Rippe, 2011). Correspondingly, the experimental curves reporting "1 – fraction unbound" can be expressed as $(1 - 1/Z)$. Next, one has to calculate the partition function, which depends on the selected model of biological interactions. Here we will use the general transfer matrix formalism, previously introduced for more complex systems including

DNA loops, multiprotein multilayer assembly and long-range interactions between DNA-bound or membrane-bound proteins (Teif, 2007; Teif, 2010).

Here, the nucleosome array is considered as a 1D lattice of binding sites with appropriate mathematical rules to make the 1D model equivalent to the corresponding 3D structure under consideration. Previously we have worked with models where the elementary lattice unit is the DNA base pair, protein amino acid or the nucleosome (Teif *et al.*, 2010; Teif and Rippe, 2010; Teif and Rippe, 2009; Teif *et al.*, 2008). Here for simplicity we will consider one half-wrap of the nucleosome as the elementary binding unit of one Swi6 dimer to one H3 tail of the nucleosome. The transfer matrices are constructed so that each matrix element $Q_n(i, j)$ contains the probabilities to find the lattice unit $n$ in a state $i$ provided the unit $n+1$ is in state $j$. Prohibited combinations of states are characterized by zero weights. The partition function is given by sequential multiplication of all transfer matrices enclosed between two unit vectors:

$$Z = \begin{pmatrix} 1 & 1 & \ldots & 1 \end{pmatrix} \times \prod_{n=1}^{N} Q_n \times \begin{pmatrix} 1 \\ 1 \\ \ldots \\ 1 \end{pmatrix} \tag{1}$$

Furthermore, we define $N$ as the number of Swi6-dimer binding sites on a nucleosome array, $c_0$ as the molar concentration of free Swi6 dimers, $K_d(n)$ as the Swi6-nucleosome dissociation constant per Swi6 dimer (the inverse value is the Swi6 binding constant $K(n)$), and $w$ as the contact cooperativity constant for interactions between neighbor Swi6 dimers (Lando and Teif, 2000; Teif *et al.*, 2002; McGhee and von Hippel, 1974). For heterogeneous binding sites of a single protein type to DNA without long-range interactions, the transfer matrix can be constructed as follows:

$$Q_{ij}(n) = \begin{pmatrix} K(n)c_0 w & K(n)c_0 \\ 1 & 1 \end{pmatrix} \tag{2}$$

Here, rows of the matrix list the states of the current lattice unit $i$, and columns list the states of the next lattice unit $j$, where only two states are allowed, bound (first row) or free (second row). The situation becomes more complicated if we recollect that the geometric distances between two Swi6 dimers located on the same nucleosome and those located on the neighboring nucleosomes are not equivalent (up to now we do not make any assumptions about the way how Swi6 proteins interact with each other). The transfer matrix formalism can account for these features. For Swi6 dimers bound to the same nucleosomes, we can consider two protein binding modes numbered as "1" and "2" characterized by the same protein concentration and binding constant. In this case each lattice unit can be in three states (bound by protein type 1, bound by protein type 2, or free), and the corresponding transfer matrix has 9 elements:

$$Q_{ij}(n) = \begin{pmatrix} K(n)c_0 & K(n)c_0 w & K(n)c_0 \\ K(n)c_0 w & K(n)c_0 & K(n)c_0 w \\ 1 & 1 & 1 \end{pmatrix} \tag{3}$$

By introducing binding constants $K_1$ and $K_2$ for different HP1 binding modes, different binding stoichiometries $m_1$ and $m_2$, and different cooperativity parameters $w_{11}$, $w_{12}$, $w_{21}$, $w_{22}$ the model can be made more complex as described previously (Teif, 2007). However, here the simple case of standard McGhee-von Hippel contact cooperativity (Eq. 2) and pair-wise cooperativity (Eq. 3) are sufficient. The difference between the latter two types of cooperativity is that according to Eq. 2 each protein can have interactions with two neighbors at saturation, while in Eq. 3 each protein can interact only with one of its neighbors which is bound to the same nucleosome.

Figure 1 illustrates this concept for Swi6 binding to mono- and di-nucleosomes. Strictly speaking, the model in Eq. 3 does not take into account that proteins #1 and #4 interact, which would require a more complicated model (see below). However, this affects only boundary conditions and thus the approximation by the matrix in Eq. 2 is justified. The only change is that the value of $w$ is squared (energy doubled) upon transition from the interaction of a protein with a single-neighbor to two-neighbors.

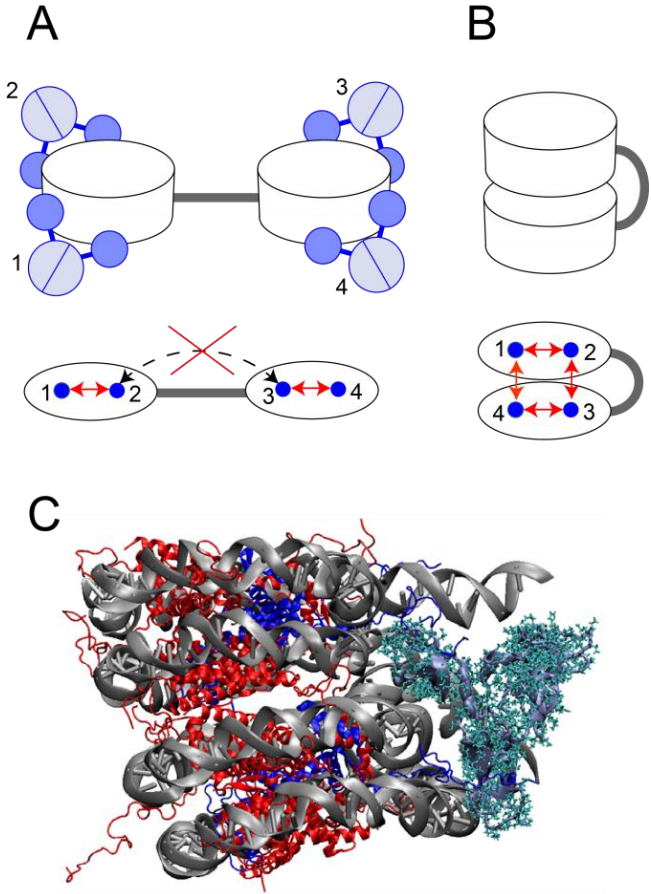

**Figure 1.** A scheme of the lattice models for different types of contact interactions between nucleosome-bound Swi6 dimers.
(A) Pair-wise cooperativity. Swi6 dimers interact only within one nucleosome. At saturation each Swi6 dimer has contact with only one neighbor.
(B) When two nucleosomes are brought into physical proximity, neighboring Swi6 dimers can interact independently of whether they are bound to the same nucleosome or to two different nucleosomes. At saturation each Swi6 dimer has contact with two neighbors.
(C) Molecular model of a HP1 dimer binding to H3 tails from two different nucleosomes in a stacked dinucleosome structure. The DNA is depicted in grey, histone H3 in dark blue, H2A, H2B, and H4 in red and the HP1 dimer in light blue.

In addition, we have considered a model in which binding is not site-specific and binding sites are independent. In terms of the molecular features it represents the case that *all* nucleosomes are equally (un)methylated and that proteins bound to neighboring sites do not interact. Under these conditions, Eqs. 1-2 can be solved analytically, and the partition function is given as

$$Z = (1 + Kc_0)^N, \qquad (4)$$

where $N$ is the number of binding sites. In a specific case when the system undergoes a conformational transition depicted in Figure 1, the partition function if given by Eq. 5 (Poland, 2001):

$$Z = (1 + Kc_0)^{N_1} + K_T(1 + Kc_0)^{N_2}, \qquad (5)$$

where *N1* and *N2* are the numbers of sites in the extended and folded dinucleosome conformation correspondingly, and $K_T$ is the equilibrium constant for the process of dinucleosome folding (Figure 1B). If dinucleosome folding does not affect the stoichiometry of HP1 binding and all binding sites remain equivalent, Eq. 5 can be simplified to

$$Z = (1 + wKc_0)^N, \tag{6}$$

where the cost of dinucleosome folding is included in *w* as the allosteric interaction between bound HP1s. The latter type of protein binding cooperativity has the same mathematical description as the McGhee-von Hippel cooperativity, but does not require direct protein-protein contacts and is induced through the conformational transition of the dinucleosome.

**Results**

As the initial reference state, we analyzed Swi6 binding to methylated and non-methylated mononucleosomes using Eqs 1-3. For the chromoshadow-domain of Swi6 a value of $K_d < 17$ nM for its self-association was determined (Canzio *et al.*, 2011). Thus, it is justified to consider Swi6 as a single dimeric species in the following theoretical analysis of its binding to different nucleosome substrates. However, it is noted that the analysis of mouse HP1β by analytical ultracentrifugation shows that the free protein has a dimerization dissociation constant of about 1 µM (Table 1) (Müller-Ott *et al.*, 2014). Accordingly, a HP1 monomer-dimer equilibrium would have to be taken into account in *in vitro* experiments conducted at protein concentrations in the range of $10^{-6}$ M and below for the mouse HP1 protein.

| Concentration (µM) | $S_1$ (S) [a] | $S_2$ (S) [a] | $M_{ave}$ (kDa) [b] |
|---|---|---|---|
| 1.1 | 2.0 ± 0.1 | – | n.d. |
| 3.5 | n.d. | n.d. | 21±1 [c] |
| 4.3 | 2.0 ± 0.1 | 2.8 ± 0.2 | n.d. |
| 7.2 | – | 2.8 ± 0.1 | n.d. |
| 10-23 µM [d] | – | 2.8 ± 0.2 | 47±1 |

**Table 1.** Concentration dependent sedimentation coefficients and molecular weights of mouse HP1β determined by analytical ultracentrifugation. Sedimentation coefficients and molecular weights of recombinant mouse His-tagged full-length HP1β were measured and analyzed at the indicated concentrations of monomer as described previously (Müller-Ott *et al.*, 2014). Values refer to $H_2O$ and a temperature of 20 ºC as standard state.
[a] Determined by sedimentation velocity ultracentrifugation and data analysis as described previously.
[b] Determined from sedimentation equilibrium ultracentrifugation at 10000 and 15000 rpm and 230 nm and 280 nm absorbance. Data were fitted to a one-component model.
[c] Sedimentation equilibrium ultracentrifugation showed systematic deviations of residues when fit to a one component model. A good fit was obtained with a two-component model with molecular weights of $M_1 = 18.8$ kDa and $M_2 = 42.6$ kDa indicative of a significant amount of HP1 monomer being present at this protein concentration.
[d] Average values from measurements at 10.4 µM and 23.4 µM which showed only a single dimeric HP1 species.

We compared two models: a simple model allowing one Swi6 dimer interacting with one H3 tail, and a more complex model, in which one Swi6 dimer interacts either with one or two H3 tails. The fit did not improve when using the more complex model. Accordingly, we used the simpler model with one mode of binding (Eq. 2). With this model, we found that Swi6 binding to mononucleosomes can be described well with a contact cooperativity parameter $w = 15$ and dissociation constant $K_d = 0.33$ μM for trimethylated and $K_d = 2$ μM for unmodified nucleosomes. This yields 1.8 kT (~1 kcal/mol) energy difference per one Swi6 dimer binding to H3K9me3 versus the unmodified H3K9 tail, which is comparable to ~2.6 kcal/mol energy of HP1 chromodomain binding to H3K9me3 peptides (Hughes *et al.*, 2007) (Figure 2).

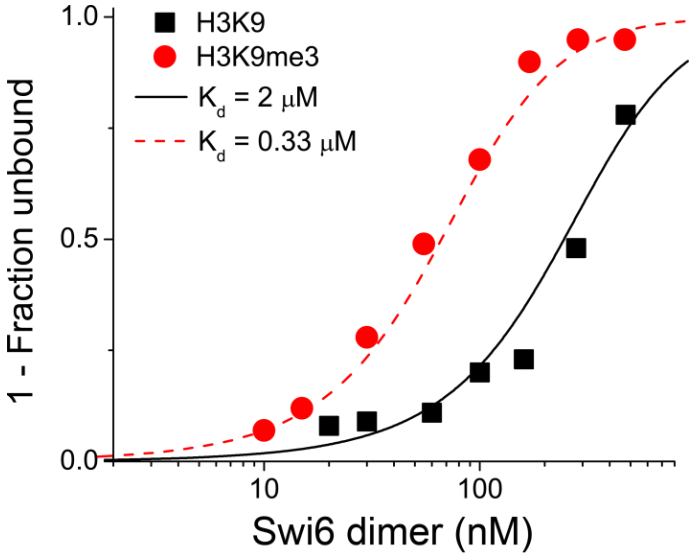

**Figure 2.** Swi6 binding to methylated (red) and non-methylated (black) mononucleosomes. Binding curves depict the fit to the data for $N = 2$, $w = 15$ according to Eq. 2, and the dissociation constant $K_d$ values indicated in the figure. Data points are from previous in vitro binding experiments (Canzio *et al.*, 2011).

Next, we analyzed the observed differences in binding to mononucleosomes and dinucleosomes with different linker lengths (Figure 3). We started with the mononucleosome set of parameters determined above. These reproduced the experimental binding curve for the 15-bp linker dinucleosome when doubling the number of mononucleosome binding sites $N$, and by additionally raising the cooperativity parameter $w$ to the second power. The molecular mechanism behind this would be that with the 15-bp linker each Swi6 dimer had one contact in a tetrameric Swi6 structure. In contrast with the 47-bp linker, two Swi6 tetramers can come close to each other in 3D so that each Swi6 dimer would interact with two, not with one neighbor dimer (see Figure 1).

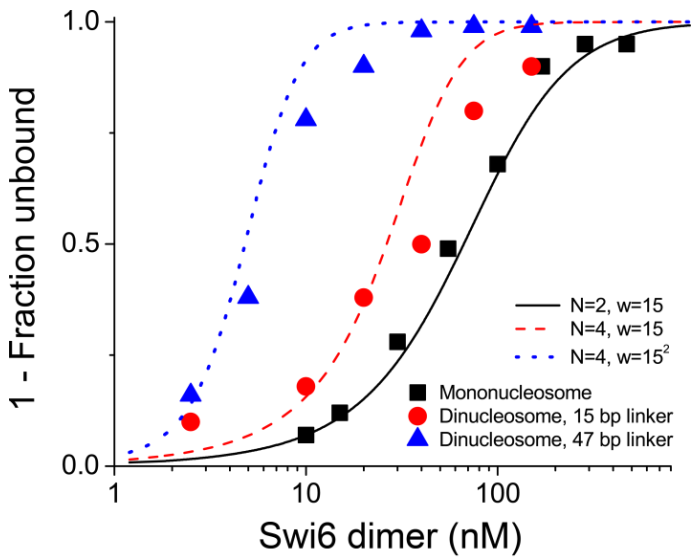

**Figure 3.** Swi6 binding to methylated mono- and dinucleosomes. Binding curves were calculated according to Eq. 1 for a value of $K_d = 0.33$ μM. Data points are from previous in vitro binding experiments (Canzio *et al.*, 2011)..

The same binding model fits for non-methylated mono- and dinucleosomes, albeit with an about 6-fold lower binding affinity (Figure 4). In this case the experimental data for the mononucleosome were fit using the algorithm in Eqs. 1-2 with the following parameters: $N = 2$, $K_d = 2$ μM, $w = 15$. By changing the stoichiometry to allow binding of twice as many proteins ($N = 4$), the dinucleosome binding curve was described well. For the dinucleosome with the short linker (15 bp) binding cooperativity was the same as in the case of the mononucleosome. For the dinucleosome with the long linker (47 bp) additional protein-protein contacts form as manifested by the two-fold increase in the protein-protein interaction energy. The corresponding cooperativity constants multiply ($w^2$ instead of $w$). The latter can be explained if long linker allows the dinucleosome to fold back so that two core particles contact each other and allow all four Swi6 dimers interact (see Figure 1).

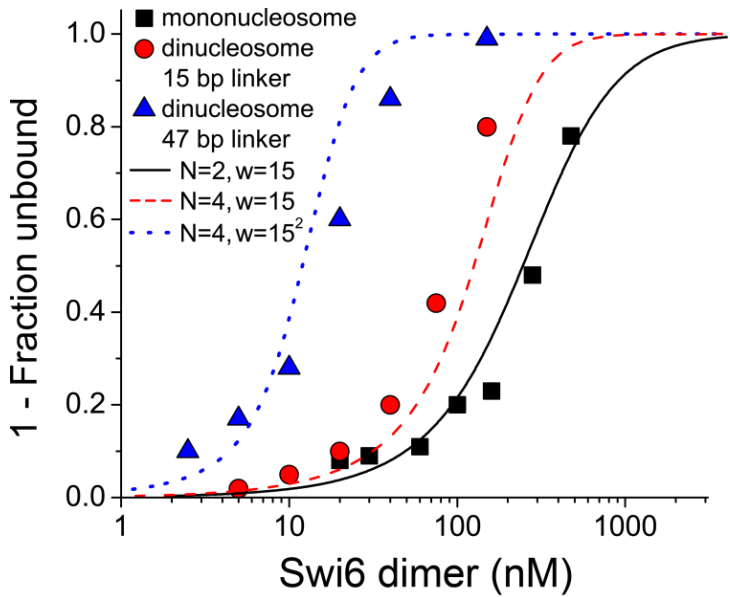

**Figure 4.** Swi6 binding to non-methylated mono- and dinucleosomes. Binding curves were calculated according to Eq. 1. $K_d = 2$ μM. Other parameters are indicated in the figure. Data points are from previous in vitro binding experiments (Canzio *et al.*, 2011).

Since data fitting in Figures 2-4 required only weak contact cooperativity, we asked whether it is possible to fit the data without contact cooperativity between HP1 dimers. Figure 5 shows the recalculation using Eq. 4, assuming that HP1 dimers do not form energetically stable complexes with each other. This fit results in higher dissociation constants ($K_d = 0.67$ μM for unmethylated and 0.17 μM for methylated mononucleosomes).

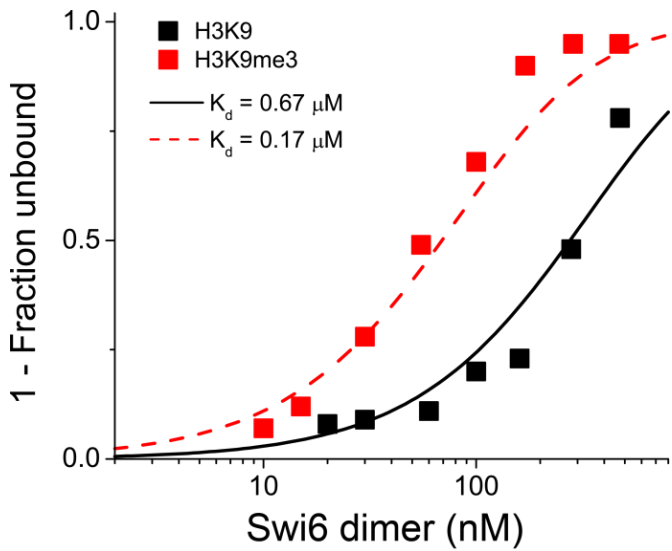

**Figure 5.** Swi6 binding to methylated (red) and non-methylated (black) mononucleosomes. Binding curves represent a theoretical fit by a non-cooperative model (Eq. 4) with $N = 2$ and dissociation constants indicated in the figure. Data points are from previous in vitro binding experiments (Canzio *et al.*, 2011).

For the dinucleosome our Monte Carlo simulations indicated a folding transition for 47-bp DNA linkers to a stacked structure (Fig. 1C) but not for 15-bp linkers. To quantitate the probability of this process, we performed a series of MC simulations of chromatin based on our previous 3D model (Kepper *et al.*, 2008). In the simulations shown in Fig. 6, the energy for the nucleosome stacking interactions at optimal distance and alignment was varied between 6 and 18 kT. As discussed previously this range of values corresponds to those derived from experiments (Kepper *et al.*, 2011).

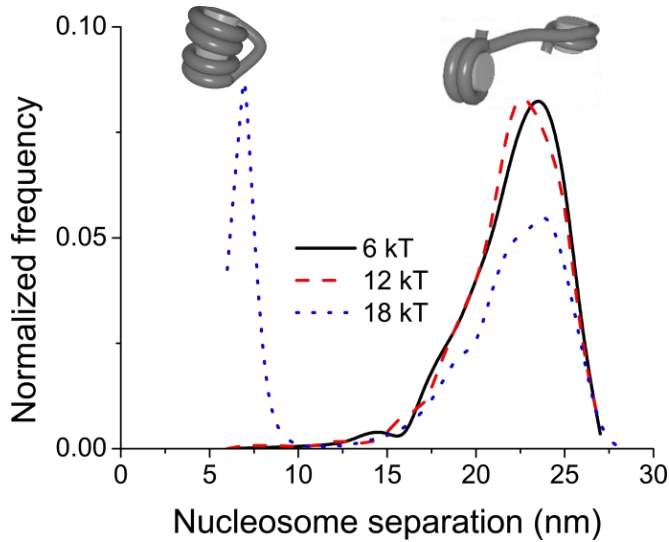

**Figure 6.** Monte Carlo simulation of the dinucleosome folding. The frequency of conformations as a function of the distance between two nucleosomes is plotted for different energies of nucleosome-nucleosome stacking interactions from 6 kT to 18 kT (values indicated on the figure). The first peak at around 6-7 nm separation distance corresponds to the folded dinucleosome conformation, where two nucleosomes are stacked on top of each other (see inset). The peak at around 23 nm separations corresponds to the stretched DNA linker between nucleosomes (linker length 47 bp).

At 12 kT the dinucleosome with 47 bp linker length showed first fold back events (distances between the centers of the two nucleosome below 11 nm). The folded back conformation was essentially absent for 15 bp linker dinucleosomes. The energy of 12 to 15 kT is comparable to the energy of two Swi6 binding to the methylated nucleosome (using $K_d$ = 0.33 µM as determined in Figures 2-4). It is noted that binding of one HP1 can stabilize the stacked nucleosome conformation and would facilitate the binding of the second dimer via an allosteric change of the nucleosome substrate. Swi6 binding curves with a dinucleosome substrate were well described with Eq. 4 (HP1 dimers do not interact) (Fig 7) and an allosteric cooperativity, $w$ = 5 (Eq. 6).

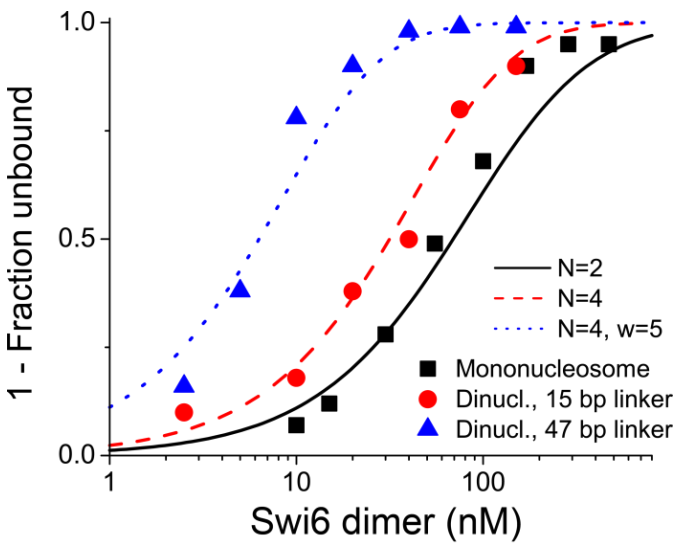

**Figure 7.** Swi6 binding to methylated mono- and dinucleosomes. Binding curves for mononucleosomes ($N$=2) and dinucleosomes with 15-bp linkers ($N$=4) were calculated according to a non-cooperative model (Eq. 4). The curve for dinucleosomes with 47-bp linkers was calculated using Eq. 6 assuming that the energy of dinucleosome folding is cast into allosteric binding cooperativity ($w$=5). $K_d$ = 0.17 µM for all curves. Data points are from previous in vitro binding experiments (Canzio *et al.*, 2011).

Finally, we analyzed Swi6 binding to the methylated 12-nucleosome arrays (Figure 8). A simple model was able to quantitatively describe the experimental data in the absence of Swi6 binding cooperativity. Binding to the 12-nucleosome array with 47-bp linkers was fit with a dissociation constant $K_d = 2$ μM and $N=24$ binding sites for Swi6 dimers (or alternatively with dissociation constant $K_d = 1.1$ μM and $N=12$ binding sites). Keeping the $K_d$ value and half the number of binding sites, we were able to fit the binding curve corresponding to the 12-nucleosomal array with 15-bp linker. In particular, if the nucleosomal array with 47-bp linkers can bind 24 Swi6 dimers (two dimers per nucleosome, as established above), the nucleosomal array with 15-bp linker can bind only 10 Swi6 dimers, which connect each second nucleosome (10 Swi6 bridges between 12 nucleosomes).

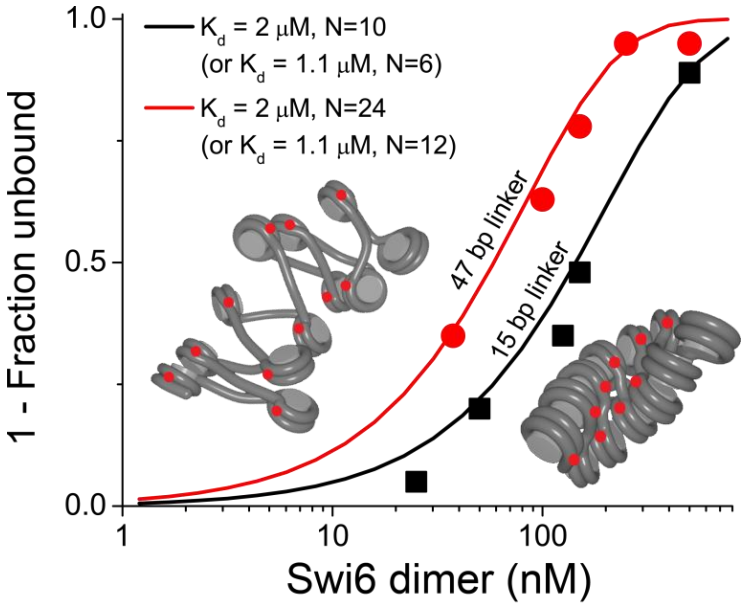

**Figure 8.** Swi6 binding to methylated arrays of 12 nucleosomes. Binding curves were calculated according to Eq. 1. Other parameters are given in the figure. Insets show representative snapshots from the Monte Carlo simulations of the 12-nucleosome array with the linker length 47 bp (left) and 15 bp (right). HP1 proteins were not included in the MC simulations of the chain, and are shown here to illustrate how the longer linker could double the number of potential binding sites with accessible H3 N-terminal tails for nucleosomes in spatial proximity. Data points are from previous in vitro binding experiments (Canzio *et al.*, 2011).

How can this model be connected to the available structural information? Our Monte Carlo simulations based on the previously described methodology (Kepper *et al.*, 2008) show that increased linker length is related to a more decondensed nucleosomal array conformation. As shown in Figure 8 (inset), regular nucleosomal arrays formed with 15 bp DNA linkers have less space for Swi6 binding in comparison with more open chromatin fibers with 47 bp linkers. Furthermore, two nucleosomes which appear to be close to each other in 3D are not necessarily two sequential neighbors in the 1D lattice, as discussed previously in the context of lattice models for chromatin (Teif and Rippe, 2010). The latter effect is demonstrated in more detail in Figure 9, where the stoichiometry of HP1-chromatin binding is justified (24 HP1 dimers for an amorphous 12-nucleosome array, versus 10 HP1 dimers for a 12-nucleosome array). Note that our MC model for long DNA linkers is fully compatible with the recent high-resolution nucleosomal array structure determined by cryo-EM (Song *et al.*, 2014). This might explain the counterintuitive experimental result that dinucleosomes bind Swi6 similar to mononucleosomes, while longer nucleosomal arrays show significant differences. The considerations above are based on a geometry of the chromatin fiber with mostly straight and cross-linked linker DNA as observed previously in a tetranucleosome crystal structure (NRL = 169 bp) without linker histones (Brasher *et al.*, 2000) and cryo-EM based structures of 12 mer arrays with linker histone H1 and NRL =177 or 187 bp (Song *et al.*, 2014). As shown here and in our previous work (Kepper *et al.*, 2008), using a local nucleosome geometry that is derived from the tetranucleosome structure will promote a more open and irregular fiber conformation for longer linker length as found in the 47-bp linker (NRL = 194) nucleosome array versus the 15-bp linker nucleosome chain (NRL = 162 bp). This conformational difference results from an increased electrostatic repulsion between negative charges in the DNA linker. Addition of linker histones would favor fiber compaction (Cherstvy and Teif, 2014) as concluded also from the appearance of folded nucleosome arrays with a conformation similar to that found in the tetranucleosome crystal structure but with longer linker length of 30 and 40 bp (Song *et al.*, 2014).

Thus, the change of the binding behavior in 15-bp vs. 47-bp linkers can be explained by a change of binding stoichiometry due to differential compaction of the fiber. In addition, the conformation of nucleosome arrays imposes additional constraints to the HP1 interactions that are not present for mono- and dinucleosomes. It should be noted that while the relative stoichiometry difference between 15-bp vs 47-bp arrays is a feature that can be robustly obtained from the analysis of the data, the absolute stoichiometry changes between mononucleosomes and arrays might be obscured by concomitant changes in binding affinities.

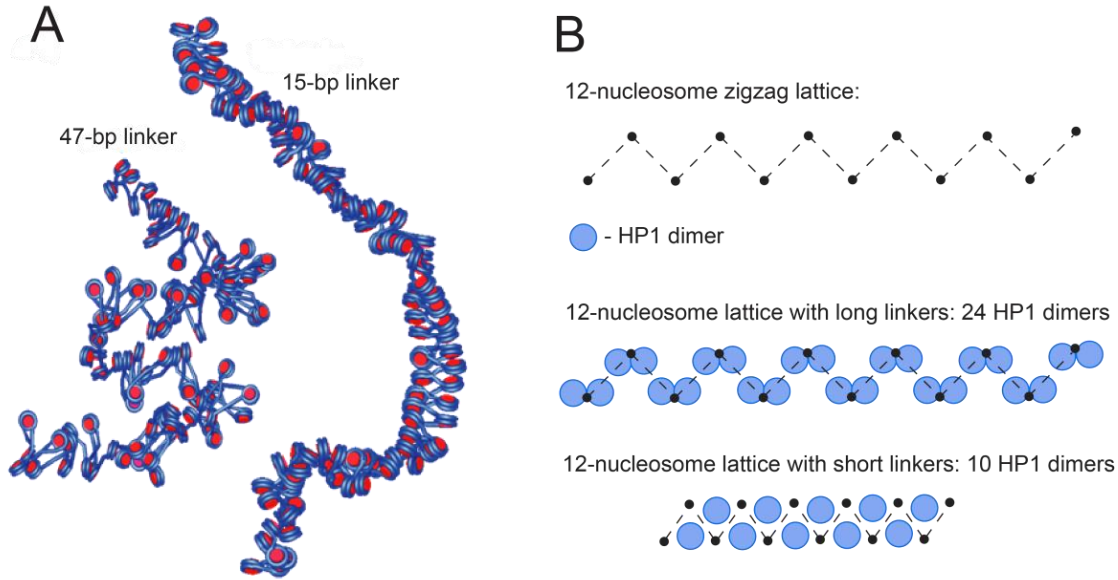

**Figure 9**. (A) MC simulation of nucleosome arrays with 15-bp and 47-bp linkers. (B) Schematic drawing explaining the potential change of the HP1 binding stoichiometry due to the geometrical constraints imposed by the shorter DNA linker length.

**Discussion**
Here, we conducted an analysis of previously published binding isotherms (Canzio *et al.*, 2011) to dissect the binding of Swi6/HP1 dimers to nucleosomal substrates. In our analysis, we confirm a several-fold increased affinity upon H3K9 trimethylation, which is compatible with other *in vitro* measurements estimates (Munari *et al.*, 2012; Jacobs *et al.*, 2001). Furthermore, we found that accounting for the different stoichiometry in the absence of cooperativity is sufficient to rationalize Swi6/HP1 binding to chromatin. Notably, the observed allosteric cooperativity of binding for dinucleosomes substrate with long but not with short DNA linkers was quantitatively consistent with stacking of nucleosomes as predicted from MC simulations. This conformation of two nucleosomes allows for simultaneous interactions of one HP1 dimer with two H3 tails as depicted in the all-atom model in Fig. 1 C. Interestingly, targeting of HP1β to heterochromatin appears to require the simultaneous recognition of two H3K9-methylated histone H3 molecules (Thiru *et al.*, 2004), which suggests that conformations of the nucleosome chain with two stacked nucleosome are preferred HP1 binding sites. The allosteric cooperativity arising due to the dinucleosome conformation change increased the Swi6 binding constant five-fold. This would be significantly larger than nonspecific interactions of Swi6 with DNA, to which an up to two-fold increase in the case of 47-bp linkers was attributed previously (Canzio *et al.*, 2011). A similar allosteric cooperativity without direct interactions between DNA-bound proteins has been described previously for the process of DNA condensation *in vitro* (Teif, 2005). Thus, we conclude that protein-protein contact cooperativity between Swi6/HP1 dimers is not needed to rationalize the available experimental binding curve. Furthermore, our recent analytical ultracentrifugation measurements of mouse HP1β proteins argue against the existence of HP1 tetramers or higher order complexes in solutions since the protein is present in a dimeric state at protein monomer concentrations up to the range of 30 µM (Table 1) (Müller-Ott *et al.*, 2014). Accordingly, cooperative interactions between chromatin-bound HP1 dimers are likely to be very weak. Furthermore, Swi6/HP1 binding to nucleosome arrays was compatible with a simple model where the stoichiometry of HP1 binding (but not the cooperativity of HP1-HP1 interaction) changes depending on the chromatin compaction state (Figures 8 and 9). Thus, a contiguous coverage of a genomic region by HP1 spreading would not be expected. This is consistent with recent experimental results which showed that about 90% of mouse HP1 proteins interact with chromatin very transiently with residence time on the second scale, while only ~10% was bound for minutes (Cheutin *et al.*, 2004; Müller *et al.*, 2009; Müller-Ott *et al.*, 2014). At the same time only a moderate enrichment of HP1 and the H3K9me3 modification for pericentric heterochromatin of 2-3 fold over other chromatin loci was observed that followed the higher DNA density in this nuclear subcompartment (Müller-Ott *et al.*, 2014). These observations are difficult to reconcile with a model in which HP1 recognition of H3K9me3 is coupled to spreading of the protein along chromatin due to protein-protein contact cooperativity. Interestingly, the Clr4/SUV39H methylases contain a chromodomain that can recognize H3K9 methylation, which could be important for the propagation of the H3K9me3 state and could compete with binding of HP1 (Al-Sady *et al.*, 2013). Thus, it will be important to further dissect, how Swi6/HP1 and Clr4/SUV39H (and possibly other factors) conspire via protein-protein interactions, recognition of the H3K9me2 and H3K9me3 modifications and long-range interactions between bound proteins along the nucleosome chain due to chromatin looping to establish, maintain and propagate heterochromatin.

**Materials and Methods**

Calculations of the binding curves were performed using the *TFnuc* software suite (Teif *et al.*, 2013; Beshnova *et al.*, 2014; Teif *et al.*, 2014; Teif, 2007) as described in the Model section.

MC simulations were performed as described previously (Kepper *et al.*, 2008). The 12-mers were simulated with maximal nucleosome interaction energy of 6 kT for the optimal stacking of two nucleosomes. The dinucleosomes were simulated with 6, 12, 18 kT maximal interaction energy. It is noted that the effective average interaction strength between two nucleosomes is significantly lower as the constraints imposed by the DNA linker partly counteract optimal stacking. The main simulation parameters are summarized in Table 2.

| Fiber Type | linker (bp) | α (°) | β (°) | γ (°) | δ (°) | ε (°) | φ (°) | c (nm) | d (nm) |
|---|---|---|---|---|---|---|---|---|---|
| cl | 15 | 70 | 140 | -70 | 20 | 0 | 0 | 3.3 | 8 |
| cl | 47 | 25 | 140 | -25 | 20 | 0 | 0 | 3.3 | 8 |

**Table 2.** Parameters of the chain of nucleosomes in the Monte Carlo simulation. The linker length, six angles and two distances are used to define the local nucleosome geometry in the Monte Carlo simulations. The angles α, β, and γ define the conformation of lowest energy in which the linker DNA is positioned relative to the nucleosome. The parameter d describes the distance of the linker DNA entering and leaving the nucleosome. The angles δ, ε, and φ define the orientation of the nucleosome relative to the DNA entry and exit points. The distance c describes the distance between the DNA entry and exit points and the center of the histone core. For details of the parameterization see Kepper *et al.*, 2008.

The all-atom model of HP1β was built using the pdb structure coordinates 1DZ1 of the chromoshadow domain (Brasher *et al.*, 2000) and 1GUW of the chromo domain (Nielsen *et al.*, 2002). The connecting hinge region was homology modeled. This HP1 structure was then attached to two stacked nucleosomes from the tetranucleosome structure (Brasher *et al.*, 2000) with their H3 tail binding to the HP1β chromodomain according to the previously reported interactions (Nielsen *et al.*, 2002). The initial structures were refined by energy minimization and molecular dynamics with the Amber 10.0 software package (Thiru *et al.*, 2004). Steric clashes of the model structures were cleaned up by energy minimization and short molecular dynamics (MD) simulations (100 ps) with a generalized Born solvent accessible surface model (ionic strength 150 mM, $\varepsilon_{internal}$ = 1.0, $\varepsilon_{external}$ = 78.5) (Tsui and Case, 2000; Hawkins *et al.*, 1996). The particle mesh Ewald method (Darden *et al.*, 1993) for non-periodic-calculations was used for the treatment of the electrostatic interactions. As convergence criterion for the energy gradient a root-mean-square of the Cartesian elements of the gradient of less than 0.05 kcal mol$^{-1}$ Å$^{-1}$ was chosen. Further molecular dynamics simulations were carried out in explicit water, at physiological salt condition (150 mM) and periodic boundary conditions in an NPT (constant pressure and constant temperature) ensemble for 5 ns.

**Acknowledgements**
V.T. acknowledges the support from the Heidelberg Center for Modeling and Simulation in the Biosciences (BIOMS) and a DKFZ Intramural Grant. Computational resources and data storage were provided via grants from the BMBF (01IG07015G, Services@MediGRID) and the German Research Foundation (DFG INST 295/27-1).